# Evaluation of Infrastructure-based Warning System on Driving Behaviors – A Roundabout Study

**Cong Zhang**
Lyles School of Civil Engineering, Purdue University
550 Stadium Mall Drive, West Lafayette, IN 47907
Email: zhan4314@purdue.edu
**Chi Tian**
School of Construction Management Technology, Purdue University
401 N. Grant Street, West Lafayette, IN 47907
Email: tian154@purdue.edu
**Tianfang Han**
Department of Psychological Sciences, Purdue University
703 3rd Street, West Lafayette, IN 47907
Email: han451@purdue.edu
**Hang Li**
School of Construction Management Technology, Purdue University
401 N. Grant Street, West Lafayette, IN 47907
Email: li4016@purdue.edu
**Yiheng Feng, Corresponding Author**
Lyles School of Civil Engineering, Purdue University
550 Stadium Mall Drive, West Lafayette, IN 47907
Email: feng333@purdue.edu
**Yunfeng Chen**
School of Construction Management Technology, Purdue University
401 N. Grant Street, West Lafayette, IN 47907
Email: chen428@purdue.edu
**Robert W. Proctor**
Department of Psychological Sciences, Purdue University
703 3rd Street, West Lafayette, IN 47907
Email: rproctor@purdue.edu
**Jiansong Zhang**
School of Construction Management Technology, Purdue University
401 N. Grant Street, West Lafayette, IN 47907
Email: zhan3062@purdue.edu

**ABSTRACT**

Smart intersections have the potential to improve road safety with sensing, communication, and edge computing technologies. Perception sensors installed at a smart intersection can monitor the traffic environment in real time and send infrastructure-based warnings to nearby travelers through V2X communication. This paper investigated how infrastructure-based warnings can influence driving behaviors and improve roundabout safety through a driving-simulator study – a challenging driving scenario for human drivers. A co-simulation platform integrating Simulation of Urban Mobility (SUMO) and Webots was developed to serve as the driving simulator. A real-world roundabout in Ann Arbor, Michigan was built in the co-simulation platform as the study area, and the merging scenarios were investigated. 36 participants were recruited and asked to navigate the roundabout under three danger levels (e.g., low, medium, high) and three collision-warning designs (e.g., no warning, warning issued 1 second in advance, warning issued 2 seconds in advance). Results indicated that advanced warnings can significantly enhance safety by minimizing potential risks compared to scenarios without warnings. Earlier warnings enabled smoother driver responses and reduced abrupt decelerations. In addition, a personalized intention prediction model was developed to predict drivers' stop-or-go decisions when the warning is displayed. Among all tested machine learning models, the XGBoost model achieved the highest prediction accuracy with a precision rate of 95.56% and a recall rate of 97.73%.

**Keywords:**

Smart Intersections; Infrastructure-Based Warning; Roundabout safety; Driving Simulator; Driver Intention Prediction

# INTRODUCTION

Traffic crashes remain a significant concern in roadway safety (1). NHTSA 2021 (2) reported that motor vehicle crashes on U.S. roadways claimed 42,939 lives in 2021, marking a 10% increase from the previous year's 39,007 fatalities. The fatality rates every 100 million vehicle miles traveled (VMT) exhibited a rise of 2.2 percent, ascending from 1.34 in 2020 to 1.37 in 2021. Notably, over 50% of all fatal and injury crashes take place at or near intersections (2).

With recent advances in sensing(3,4), communication(5,6), and edge computing technologies(7), the concept of smart intersection (8) is proposed to reduce fatalities, injuries, and improve mobility. Perception sensors (e.g., cameras, lidars, and radars) installed at a smart intersection play a vital role in monitoring the traffic environment while edge computing devices take the data and make real-time decisions. One notable advantage of smart intersections is their ability to eliminate visual obstructions commonly experienced by vehicle onboard sensors and provide better situation awareness. In addition, not only can smart intersections obtain the current traffic condition, but they can also make predictions of future traffic states (e.g., vehicle trajectories), which is crucial for accident prevention and traffic flow optimization.

Through vehicle-to-everything (V2X) communication, information or warnings from the smart intersection can be delivered to make informed decisions, reducing the risk of accidents, and improving overall road safety. The Cooperative Intersection Collision Avoidance Systems Initiative (CICAS) in the U.S. emphasizes the importance of real-time warnings in vehicles and infrastructure to effectively prevent crashes (9). V2X communications enable approaching vehicles to receive crucial data on traffic conditions, signal timings, and potential hazards. By combining in-vehicle warnings and infrastructure-based notifications, drivers gain the information needed to make decisions and take appropriate actions.

One example of the smart intersection (10) was recently constructed at the roundabout of State St. and W. Ellsworth Rd. in Ann Arbor, Michigan. This intersection integrates a roadside cooperative perception system with an edge-cloud structure and multiple sensors. Utilizing high-performance detection algorithms on edge devices, the system efficiently analyzes data on the cloud. Its capabilities extend to various applications, including traffic data collection and road safety studies, with a particular emphasis on infrastructure-vehicle cooperative perception.

Extensive research has been conducted on drivers' behaviors with V2I or V2V communication-based warning/information systems. One of the early works developed a driver smart assistance system (DSAS) via radio frequency identification (RFID) technology, aimed at enhancing traffic safety in work zones through real-time communication between vehicles and the roadside (11). This system could provide verbal and visual warning messages to drivers as they approached work zones. The study examined the impact of 20 participants on vehicle speeds, drivers' responses, traffic safety, and vehicle emissions. The results showed that DSAS facilitated earlier driver action, leading to deceleration and reduced speeds. Horberry et al. developed a human-centered multi-modal warning system that incorporates visual, auditory, and tactile elements to address driver fatigue and distractions, aiming to reduce crashes (12). Reinmueller et al. conducted a driving-simulator study to evaluate the impact of adaptive

Forward Collision Warning (FCW) on drivers' behavior (13). By adjusting the warning threshold based on driver distraction, the study found that adaptive FCW leads to adverse behavioral adaptation, which includes shorter time headways and Time-to-Collision (TTC) values in longitudinal control. Butakov and Ioannou introduced a personalized pace optimization algorithm designed for vehicles equipped with the Advanced Driver Assistance System (ADAS) traveling through signalized intersections (14). By considering individual driver preferences and characteristics, the method determined an optimal driving pace specific to each driver. This approach resulted in improved fuel economy, reduced waiting time, and addressed the driver's preferences.

Roundabout is a challenging scenario for human drivers. The most common crashes at multilane roundabouts are the "merge-type" crash and left turn (15). The Road Commission of Washtenaw County, MI, USA, conducted an online survey in April 2016 to study drivers' perception of roundabouts, focusing on merge errors and left turns (Roundabout User Questionnaire, 2016) (16). Out of 4,300 respondents, 34% mistakenly believed that entering a roundabout is a merging maneuver and ignore yield-to-circulating traffic, which may cause potential crashes when the vehicle enters the roundabout at an inappropriate time. The smart intersection implementation can improve roundabout safety by monitoring the vehicle entering maneuvers and generating warnings to the vehicles within the roundabout through V2X communication when improper merging behaviors are observed or predicted.

There is limited research exploring infrastructure-based warning systems on drivers' behaviors in roundabout scenarios. In this study, we conducted a driving-simulator study to investigate how warnings can effectively influence driving behaviors and improve roundabout safety. Instead of using commercial simulators, we developed a customized driving simulator that integrates SUMO and Webots. This platform offers a combination of traffic-level and experiment-level simulations, allowing for more comprehensive interactions and analysis between the driver and background traffic. A real-world roundabout in Ann Arbor, Michigan was selected as the test site, because of its complex geometry and high crash numbers. We simulated the driving scenario where an aggressive vehicle enters the roundabout improperly, producing an interaction with the ego vehicle controlled by a driver. We assumed the infrastructure is monitoring the behavior of the aggressive vehicle and sending warning messages when an improper merging is detected. The warning message was sent to the ego vehicle through V2X and displayed on the vehicle's dashboard. We recruited 36 participants to navigate the roundabout and analyzed their driving behaviors under various warning conditions and different aggressive-driving levels. Additionally, we developed a personalized intent prediction model to predict drivers' passing or stopping decisions. The results indicated that the warning information effectively increases driver safety, with earlier warnings providing greater driving comfort and reduced crash probability. The personalized prediction model can accurately predict the driver's stop-or-go intent when the warning is issued. The main contributions of this study are:

(1) A novel customized driving simulator that integrates SUMO and Webots was developed.

(2) The impacts of an infrastructure-based warning system on driving behaviors at roundabout driving scenarios under various aggressive and warning conditions were tested and evaluated.
(3) The intent prediction model considered personalized information including driving behaviors, traffic safety attributes, and eye tracker characteristics, and achieved high-accuracy prediction results.

The rest of the paper is organized as follows. Development of the driving-simulator platform is introduced in section 2. The driving-simulator study is presented in section 3 and the corresponding data analyses in section 4. In section 5, the intent prediction model is introduced to predict drivers' stop-or-go decisions. Finally, conclusions and future directions are exhibited in section 6.

## CO-SIMULATION PLATFORM DEVELOPMENT

Existing commercial driving simulators (e.g., STISIM, VirtualDriver, DriveSafety)(17) exhibit certain limitations, such as predefined road geometry and areas, pre-programmed vehicle states, and lack of capability to capture realistic vehicle behaviors. These limitations motivated us to develop a co-simulation platform that combines microscopic traffic simulation (SUMO) and a robotic simulator (Webots), with capabilities to be extended to include autonomous driving functions (e.g., perception sensors).

**Figure 1** shows a snapshot of the co-simulation platform. The Webots view represents the simulation environment seen by the driver including the ego vehicle (controlled by the human driver through the Logitech Racing Wheel in **Figure 1(H)**), roadways, lane markings, and traffic signs (**Figure 1(E)**). Vehicle speed (**Figure 1(D)**) and warning information (when available) (**Figure 1(C)**) are also displayed in the ego vehicle. Additionally, drivers can utilize the rear mirror to assess the state of the vehicles behind them. In the SUMO view, the green vehicle is the representation of the ego vehicle in Webots while red vehicles are background traffic. Vehicle states in SUMO and Webots are synchronized. A GP3 HD eye tracker manufactured by GazePoint is connected to the system (**Figure 1(I)** to capture the eye movements.

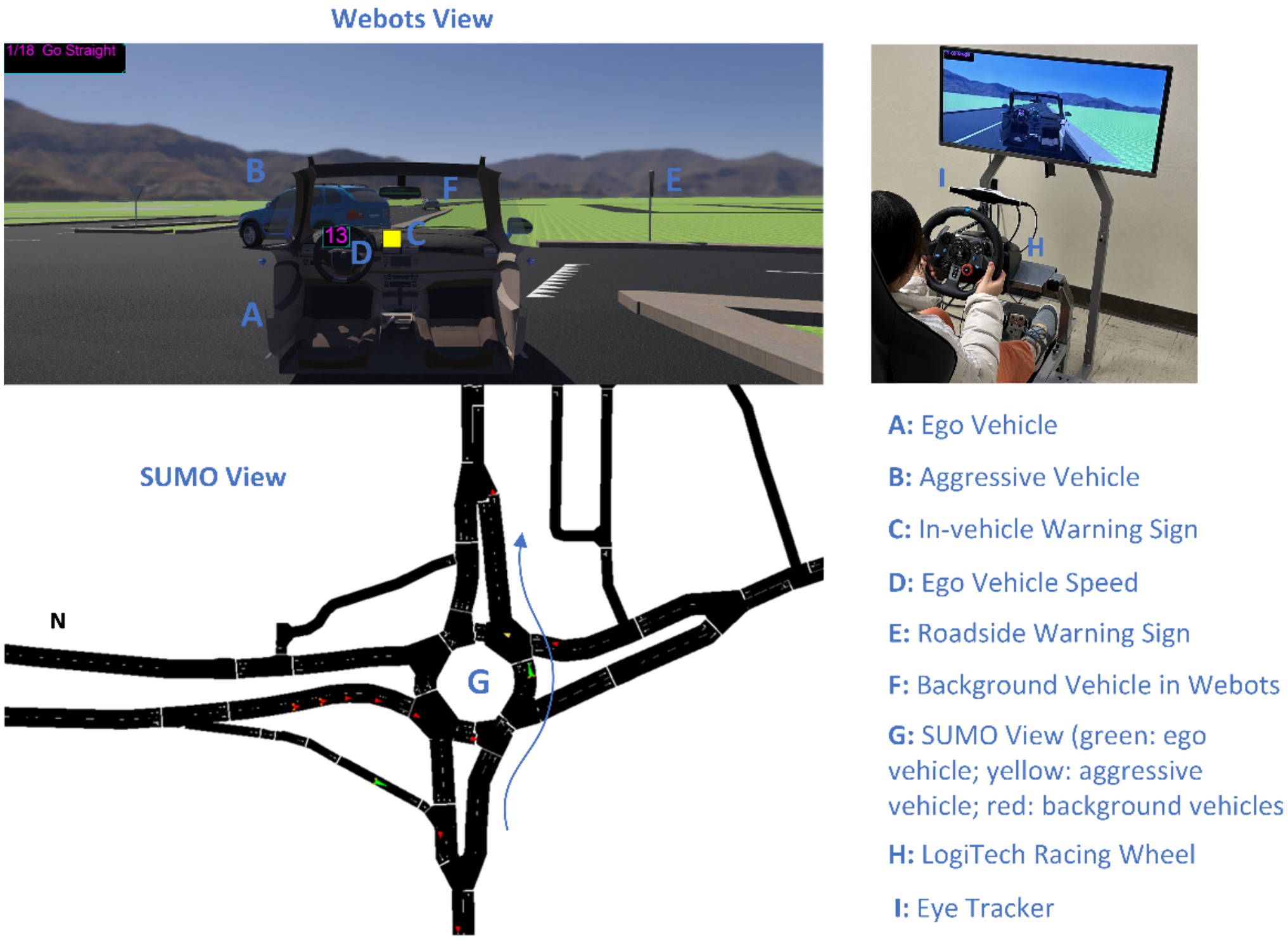

**Figure 1** Multi-level Co-Simulation Platform

# METHODS

## Scenario

The roundabout at Ellsworth and State St, Ann Arbor, MI was selected as the study site because of its high crash rate (a total number of 534 crashes from 2018-2022)(18), complex geometries, availability of real-world traffic data (e.g., Micy 2.0 dataset (19)), and expert and public opinions from our survey (20) and interviews(21). Through crash report analysis, we identified that the most frequent crash type at this roundabout is "failed to yield" (53%). It meant that a vehicle outside the roundabout tried to enter with an improper gap and caused a collision with a vehicle within the roundabout. Therefore, we decided to construct the merging scenario as the case study to test the effectiveness of the infrastructure-based warning.

The geometry of the roundabout was extracted from OpenStreetMap and calibrated in both SUMO and Webots. Traffic signs (e.g., speed limits, yield signs, roundabout directional signs) were added to Webots to mimic real-world environments. In the experiment, participants were instructed to drive the simulator (i.e., ego vehicle in **Figure 1(A)**) and completed trips navigating through the roundabout from West to East. During each trip, the driver interacted with an aggressive vehicle (i.e., the blue vehicle in **Figure 1(B)**) traveling through the roundabout from South to North. The aggressive vehicle had three aggressive levels (e.g., low, medium, and high). The low level meant that the aggressive vehicle followed traffic rules and yielded to the ego vehicle at the roundabout entrance. For medium and high levels, the aggressive vehicle entered the roundabout improperly with predefined time headways to the ego vehicle. We adopted two distinct time headways (i.e., 1.5 seconds for medium aggressive level and 0.5 seconds for high aggressive level) to represent different danger levels. During the experiments, vehicle dynamics (e.g., location, speed, and acceleration profiles of the two vehicles) and driving behaviors (e.g., throttle, brake, steering wheel angle) were recorded by the driving simulator, and the eye movements were captured by the eye tracker (**Figure 1 (I)**). A countermeasure was designed based on the assumption that the infrastructure is equipped with perception sensors to detect real-time traffic and monitor safety-critical events. In the roundabout merging scenario, the infrastructure can obtain real-time locations and speeds of both vehicles within (i.e., ego vehicle) and approaching (i.e., aggressive vehicle) the roundabout. Meanwhile, we assumed that both the infrastructure and the ego vehicle were equipped with V2X devices, enabling them to communicate in real time. As a result, a warning message was generated from the infrastructure and transmitted to the ego vehicle and displayed in the vehicle (**Figure 1 (C)**), indicating a potential collision as the aggressive vehicle fails to yield. Based on the trajectory prediction, assuming two vehicles traveling at constant speeds, the infrastructure sent the warning message 1 second or 2 seconds before the potential collision time.

**Subjects**

Thirty-six participants (18 females, 18 males) between the ages of 18 and 29 years (*M*=19.75, *SD*=1.96) were recruited from SONA and LinkedIn. SONA is an online participant recruitment system by the Psychology Department at Purdue University to engage undergraduate students in research studies. Participants had an average driving experience of 3.36 years (*SD*=1.27) and had to possess a valid U.S. driving license, be at least 18 years of age, and have normal vision or corrected-normal vision with lenses. Furthermore, they should not experience motion sickness or dizziness. The study was approved by Purdue University's Purdue's Institutional Review Board (IRB).

**Procedures**

The entire experiment took approximately 60 minutes to complete. First of all, all participants needed to read, understand, agree, and then sign a consent form. Then, a pre-survey was conducted to collect their demographic information (e.g., age, gender, education, and driving experience). Before the formal experiment, all participants engaged in several warm-up driving exercises to familiarize themselves with the driving simulator. Following this, the eye tracker was calibrated for each participant to ensure accurate data collection. Each participant was asked to complete 18 trips, including three aggressive levels (low, medium, high), three warning countermeasures (no warning, warning issued 1 second in advance, warning issued 2 seconds in advance), and repeated 2 times to capture the variations in driving behaviors, resulting in a 3× 3× 2 experiment design. The order of trips was randomly assigned for each participant. The formal experiment consisted of three sections, with 6 trips in each section and 5 minutes break between each section. When driving, the participants were asked to follow the speed limit of the road (45 mph) and within the roundabout (15 mph) as closely as possible. After each trip, participants were asked a verbal question regarding their risk perception: *"When I was driving at the roundabout, I believe the risk of getting into an accident was?"*. They were provided with 5 response options: extremely low, somewhat low, neither low nor high, somewhat high, and extremely high. Then, the participants were required to complete a post-survey for system evaluation of future research. On fulfilling the requirements of the experiment, participants received 2 credits in the course as compensation. Note that one trial of a participant was removed from the data analysis due to an incorrect traveling route.

# DATA ANALYSIS

## Data Processing

Three types of data were collected during the experiment, including vehicle trajectory data (e.g., speed, acceleration), eye tracker data (e.g., eye fixation duration, pupil diameter), and survey data (e.g., demographic, risk perception). Vehicle trajectories were collected with a resolution of 10 Hz. To reduce the measurement noise, trajectories were first smoothed by the Kalman filter, as shown in **Figure 2**.

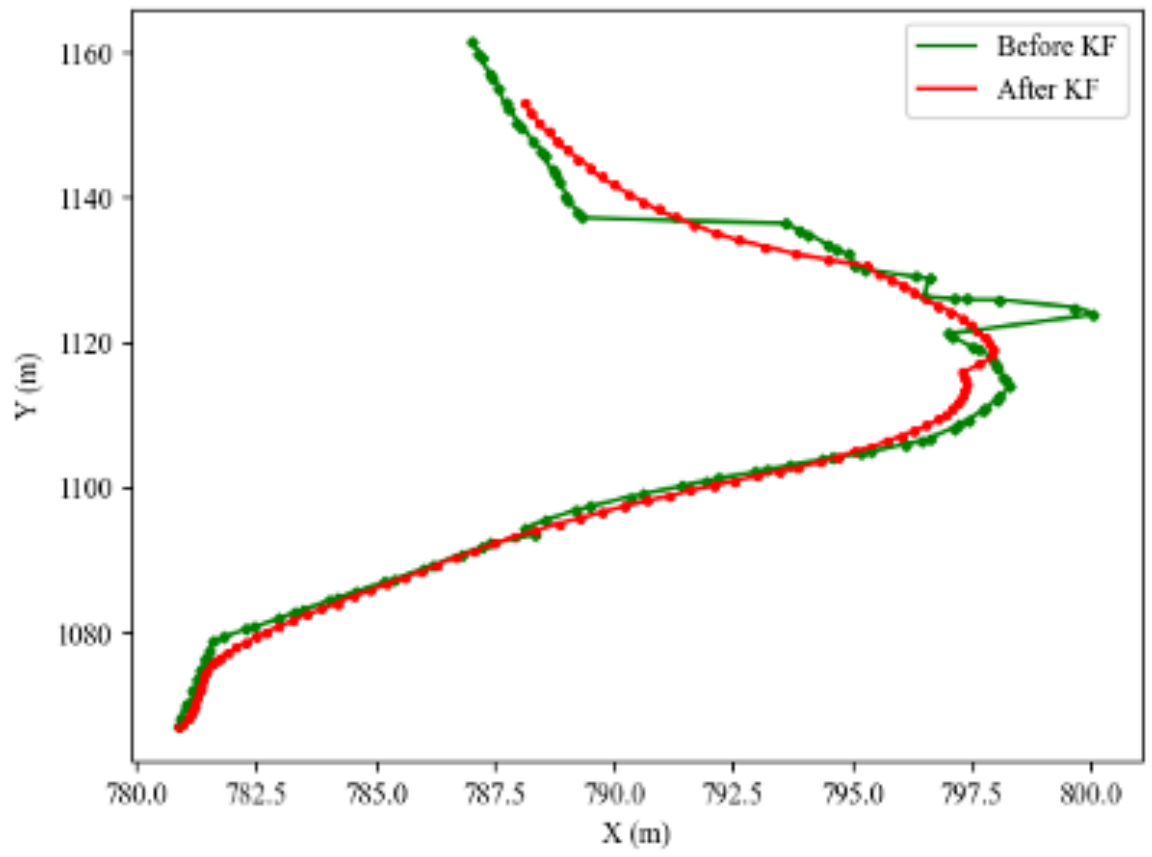

**Figure 2** Trajectory before and after Kalman filter and speed profile after Kalman filter

Four steps were conducted to process eye-movement data. First, the area of interest (AOI) is labeled. In this study, four AOIs were considered: the aggressive vehicle, speed information, warning information, and the road ahead. Second, the timestamps such as when the driver entered and exited the roundabout were labeled. Third, the eye-fixation files containing fixation duration, pupil diameter timestamp, and AOI fixation were exported from the GazePoint Analysis software. Fourth, a Python script was developed to extract and calculate eye movement indicators, such as fixation duration, number of fixations, and average fixation duration on each AOI. In addition, pupil diameter-related indicators were calculated, such as minimum, maximum, and mean pupil diameters.

Finally, the risk perception survey data were processed. **Figure 3** shows the box plot of the risk perception in the three aggressiveness levels (e.g., low, medium, high). The red line is the median and the dot is an outlier. It can be seen that as the aggressive level increases, drivers'

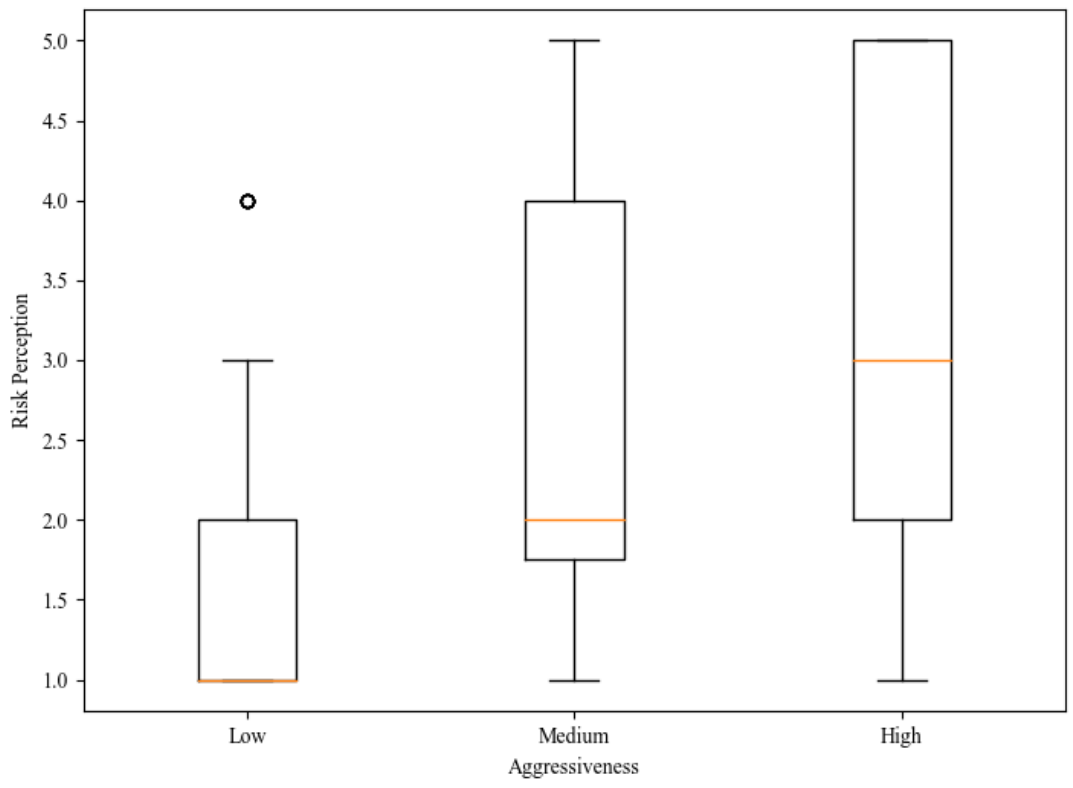

risk perception also increased.

**Figure 3** The box plot of the risk perception under different aggressiveness levels

## Metrics

To evaluate the driving behavior and cognitive load under different scenarios as well as associated impacts on safety and driving comfort, the following metrics were calculated based on the collected data.

*Time-to-collision (TTC)*

TTC has been widely applied as a safety indicator. At an instant $t$, TTC is defined as *the time that remains until a collision between two vehicles would have occurred if the collision course and speed difference are maintained* (22). **Figure 4** shows the schematic of TTC analysis with two angled approaching vehicles $i$ and $j$, representing the ego and aggressive vehicles in our study. Based on the definition, the TTC is calculated as:

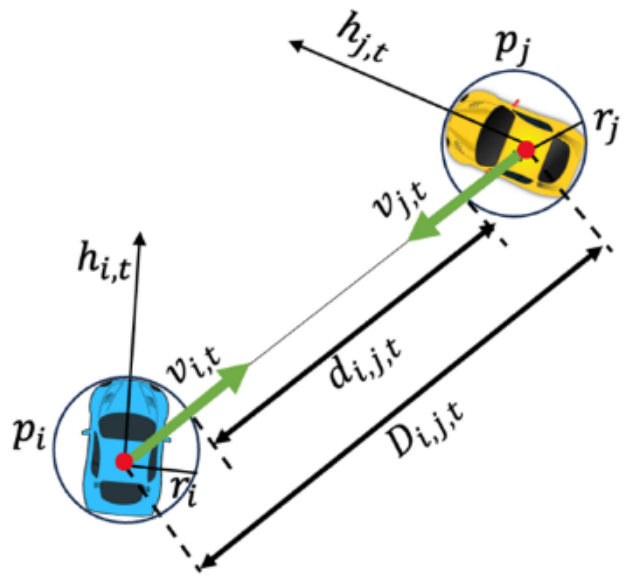

**Figure 4** The schematic of TTC analysis

$$TTC = \frac{D_{i,j,t} - r_i - r_j}{v_{i,t} + v_{j,t}} = \frac{d_{i,j,t}}{v_{i,t} + v_{j,t}} \quad (1)$$

Where, $h_{i,t}$ and $h_{j,t}$ are the headings of the two vehicles; $v_{i,t}$ and $v_{j,t}$ are the crossing direction speeds; $d_{i,j,t}$ is the net distance between two vehicles considering the physical radii $r_i$ and $r_j$, and $t$ is the time instant.

**Figure 5** presents an example of calculated TTC from the same participant under different scenarios. **Figure 5(a) and Figure 5(b)** describe the TTC under the medium and high aggressive levels with 1 second warning in advance. Similarly, **Figure 5(c)** and **Figure 5(d)** depict TTCs under medium and high aggressive levels, with 2 seconds warning in advance. It can be seen that for the same driver when only 1 second advanced warning is provided, higher aggressive level results in a much lower TTC (a crash in this case). When 2 second advanced warning was provided, the minimum TTCs between the medium aggressive level and high aggressive level were similar. This suggests that a 1-second advanced warning may not provide enough time for the driver to react to aggressive driving behavior, which was consistent with the overall data analysis presented in the next section.

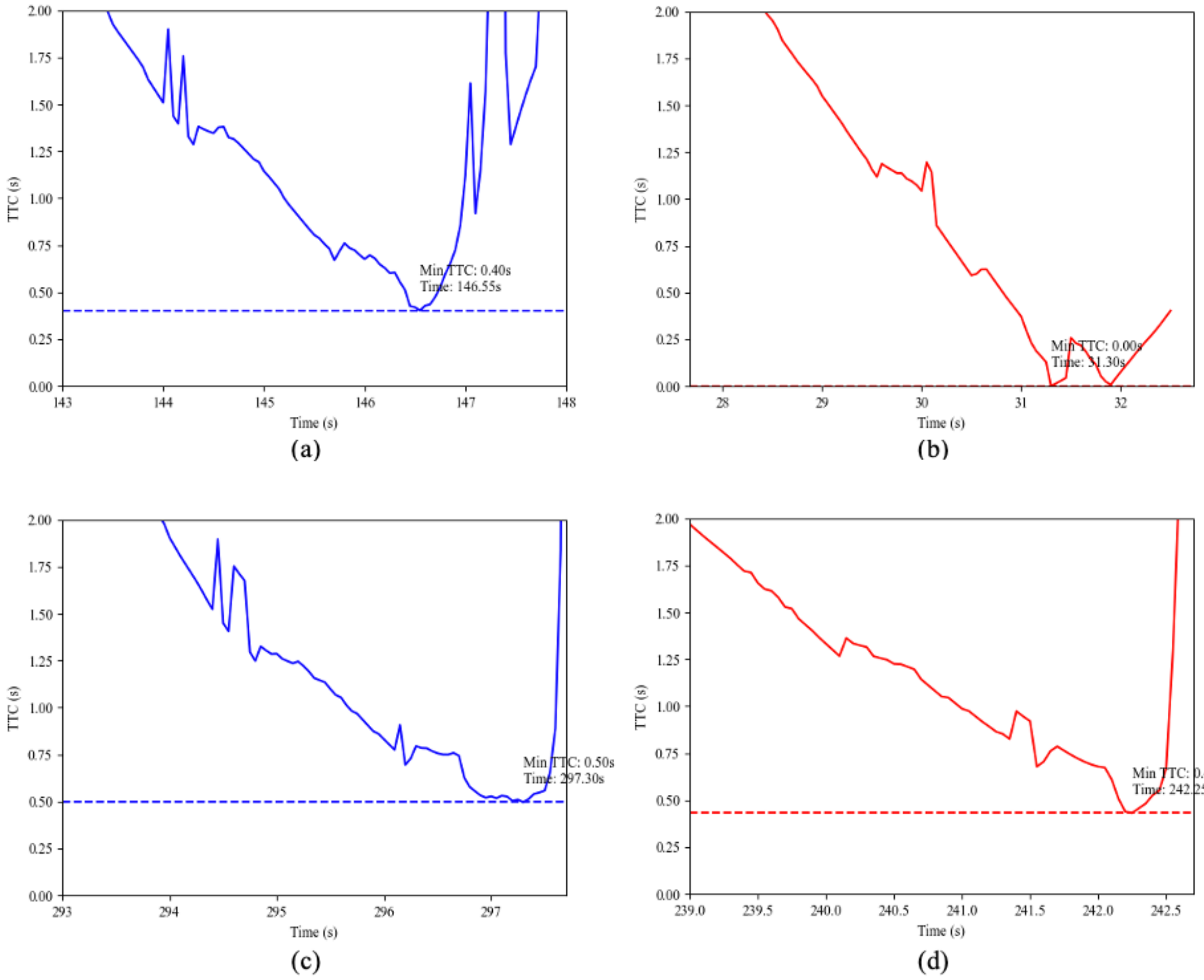

**Figure 5** Time-to-collision comparison under different aggressive levels

*Time headway* ($h_t$)

As a crucial indicator for assessing the criticality of a traffic situation, time headway represents *"the elapsed time between the front of the lead vehicle passing a specific point on the roadway and the front of the following vehicle passing the same point"(23).* The headway to the possible collision point is defined as:

$$h_t = \frac{D_{j,t}}{V_{j,t}} - \frac{D_{i,t}}{V_{i,t}} \quad (2)$$

Where, $D_{i,t}$ and $D_{j,t}$ are the distances of two vehicles to the potential collision location. $V_{i,t}$ and $V_{j,t}$ are the speeds of the two vehicles.

*Crash potential index* (*CPI*)

Cunto and Saccomanno (2008) proposed the CPI as a safety metric (24), which considers the varying braking capabilities of different vehicles and different road conditions over time. The CPI for vehicle $i$ is defined as:

$$CPI_i = \frac{\sum_{t=t_i^e}^{t_i^f} P(DRAC_{i,t} > MADR_{i,t})\Delta t b}{T_i} \tag{3}$$

Where, $t_i^e$ is defined as the time when the ego vehicle enters the roundabout; $t_i^f$ is the time when the ego vehicle arrives the possible collision location; $\Delta t$ is the observation time interval; $b$ is a binary variable, and equals 1 if $DRAC_{i,t}$>0, and 0 otherwise; $T_i$ is the total simulation time for vehicle *i*. $T_i$ used in the current study is calculated from the time point when the vehicle enters the roundabout to the minimum TTC time point.

The relevance of the deceleration rate to avoid the crash (DRAC) is a safety performance measure, which considers both speed differentials and decelerations under dangerous scenarios. In this study, $DRAC_{i,t}$ reflects the required deceleration of vehicle $i$ to avoid a crash at time $t$:

$$DRAC_{i,t} = \frac{(V_{i,t} - V_{j,t})^2}{2d_{i,j,t}} \tag{4}$$

In $P(DRAC_{i,t} > MADR_{i,t})$, $P$ is the probability that DRAC is greater than MADR for vehicle *i* at time *t*. It presents the probability of a vehicle's DRAC surpassing its maximum available deceleration rate (MADR) or braking capability at every time step (24). MADR is a metric measuring the braking ability, which varies based on vehicle characteristics (e.g., weight, tire, braking ability) and road surface conditions (e.g., dry, wet). MADR is assumed to follow a truncated normal distribution as in (24).

*Acceleration Noise (AN)*

The concept of acceleration noise (AN) was initially introduced by (25), which is the root mean square error of the vehicle's acceleration (17,18). AN for vehicle *i* is defined using the following equation.

$$AN_i = \sqrt{\frac{\sum_{t=t_i^e}^{t_i^f}(a_{i,t} - \bar{a}_i)^2}{T_i}} \tag{5}$$

Where $a_{i,t}$ is the acceleration for vehicle *i* at time *t*, $\bar{a}_i$is the average acceleration for vehicle *i* during the study time interval $T_i$.

*Pupil Diameter (PD)*

PD is a critical metric to evaluate the cognitive load through the driver's visual attention (27). The minimum pupil diameter measures the minimum eye pupil diameter when the ego vehicle is within the roundabout. Since the eye tracker captures both right and left eye pupils, the average minimum pupil diameter is calculated by equation (6):

$$\overline{PD} = \frac{(PD_l + PD_r)}{2} \tag{6}$$

Where, $\overline{PD}$ is the average minimum pupil diameter, $PD_l$ and $PD_r$ are the minimum left pupil diameter and right pupil diameter measured in pixels.

*Fixation Duration*

The fixation duration of drivers has been identified as a significant feature for the prediction and recognition of driving maneuvers (28) and is defined as:

$$MFD_{road} = \frac{FD_{road}}{N_{road}} \tag{7}$$

Where, $MFD_{road}$ is the mean fixation duration on the road, $FD_{road}$ is the total fixation duration on the road. $N_{road}$ is the number of fixations on the road, which measures how many times the drivers fixate on the road.

**Data Analysis**

To explore whether the infrastructure-based advanced warnings have an impact on driving behavior and improve safety, we conducted statistical analyses on the collected data using SPSS.

We used minimum TTC and CPI, and maximum DRAC as safety indicators because the first two provide valuable information about the urgency of taking evasive actions and the proximity of potential collisions, and the last one represents the maximum deceleration achievable to avoid a collision. The CPI was computed from the time the ego vehicle entered the roundabout to the minimum TTC time point. Analyses of variance (ANOVAs) were conducted on TTC, CPI, and DRAC, as shown in **Table 1**. A main effect of aggressiveness is shown for all variables, validating the aggressive driving manipulation. The interaction effects of warning and aggressiveness on CPI and DRAC were statistically significant (see the third row for each metric in **Table 1**). The overall results indicated that early warnings alleviated the impact of aggressive driving behaviors on objective safety evaluations (CPI, DRAC). By providing drivers with increased time to react and the ability to maintain a safer distance from potential hazards, the warning system effectively enhanced their overall safety performance. These results emphasized the effectiveness of the infrastructure-based warning system in improving driving safety and reducing the likelihood of accidents in the roundabout merging scenario.

**Table 1 Results of ANOVA on risk perception to investigate the effect of warning and aggressiveness.**

| Metrics | factors | df | Mean Square | F | *p* | Partial Eta Squared |
|---|---|---|---|---|---|---|
| TTC | warning | 2 | 1.547 | 2.665 | 0.077 | 0.071 |
| | aggressiveness | 2 | 2.356 | 5.241 | 0.008 | 0.13 |
| | warning * aggressiveness | 4 | 0.072 | 0.172 | 0.953 | 0.005 |
| CPI | warning | 2 | 0.064 | 9.072 | <.001 | 0.206 |
| | aggressiveness | 2 | 0.3 | 27.248 | <.001 | 0.438 |
| | warning * aggressiveness | 4 | 0.025 | 4.864 | 0.001 | 0.122 |
| DRAC | warning | 2 | 167.752 | 9.126 | <.001 | 0.207 |

| | | | | | |
|---|---|---|---|---|---|
| aggressiveness | 2 | 842.653 | 31.974 | <.001 | 0.477 |
| warning * aggressiveness | 4 | 40.47 | 3.024 | 0.02 | 0.08 |

From the warning side, a warning issued 2 seconds in advance led to higher TTC than the other two conditions ($F(1, 35) = 4.05$, $p = .049$), as shown in **Figure 6(a)**. This implied that a warning issued 2 seconds in advance improved driving safety compared to no warning conditions or a warning issued 1 second in advance. Additionally, the scenarios without warning resulted in the highest CPI ($F(1, 35)=8.656$, $p<.006$), but there was no significant difference in CPI when the warning was issued 1 second or 2 seconds in advance, as shown in **Figure 6(c)**. To explore the difference, we further investigated the maximum deceleration of warning issued 1 second and 2 seconds in advance. The maximum deceleration was measured from the moment the ego vehicle entered the roundabout until reaching the point of minimum TTC. The results showed that a warning issued 2 seconds in advance resulted in a lower maximum deceleration compared to the warning issued 1 second in advance ($F(1,35)=7.391$, $p=.01$). This finding indicated that although the safety impacts (i.e., CPI) of different warning times are similar, earlier warning enabled drivers to react more comfortably within the extended timeframe. Moreover, note that the warning conditions (combining the 1-second and 2-second warnings) demonstrated a lower maximum DRAC ($F(1,35)=12.354$, $p=.001$) in comparison to the no-warning condition (as shown in **Figure 6(e)**). This observation further reinforced the evidence that providing in-vehicle warnings could significantly reduce the maximum deceleration needed to avoid crashes. In conclusion, these results implied that an in-vehicle warning could improve safety and driving comfort, as long as the warning was provided sufficiently ahead of the hazard (e.g., 2 seconds in our study).

To better illustrate the effectiveness of the advanced warning, an example of speed and acceleration profiles is shown in **Figure 7** under the high aggressive level from the same participant. When provided with the 2-second warning in advance, the acceleration profile showed a much smoother response with a maximum deceleration of around -7.8 $m/s^2$ (**Figure 7(c)**) (average deceleration = -2.76 $m/s^2$) during the merging event, compared to a maximum deceleration of -9.8 $m/s^2$ (**Figure 7(b)**) (average deceleration = -6.11 $m/s^2$) with 1-second warning in advance, and -10.5 $m/s^2$ (**Figure 7(a)**) (average deceleration = -7.098 $m/s^2$) without warning. Furthermore, analyzing the speed profiles revealed that without warning, the driver experienced the shortest brake duration of less than 1 second (**Figure 7(a)**). However, with the 1-second warning in advance, the braking duration increased to 1.5 seconds (**Figure 7(b)**) and 3.5 seconds under the 2-second warning (**Figure 7(c)**). These findings indicated that the warning system effectively enhanced driving comfort by facilitating smoother vehicle speed control and avoiding harsh deceleration.

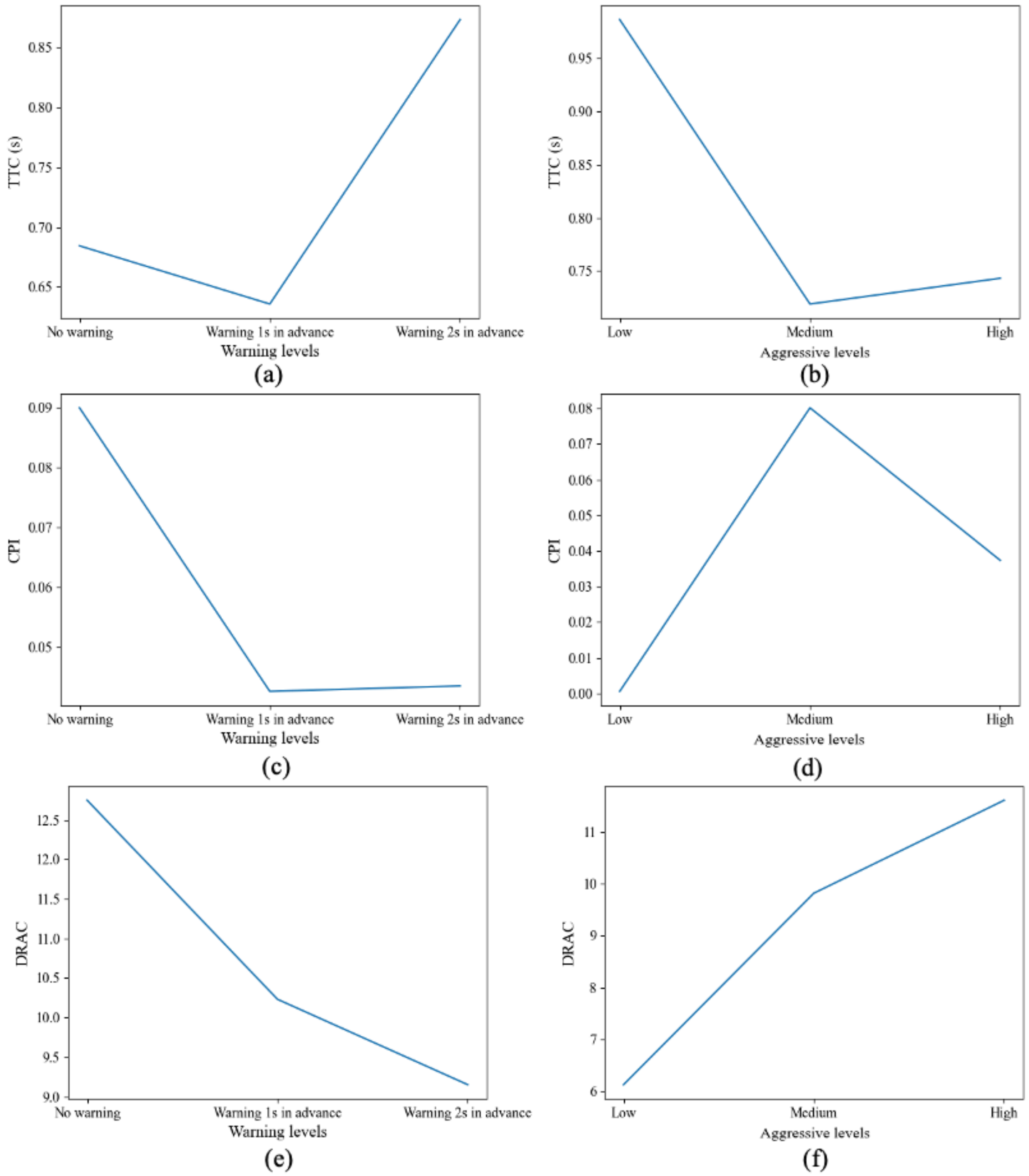

**Figure 6** The effects of warning levels on TTC (a), the effects of aggressive levels on TTC (b), the effects of warning levels on CPI (c), the effects of aggressive levels on CPI (d), the effects of warning levels on DRAC (e), the effects of aggressive levels on DRAC (f)

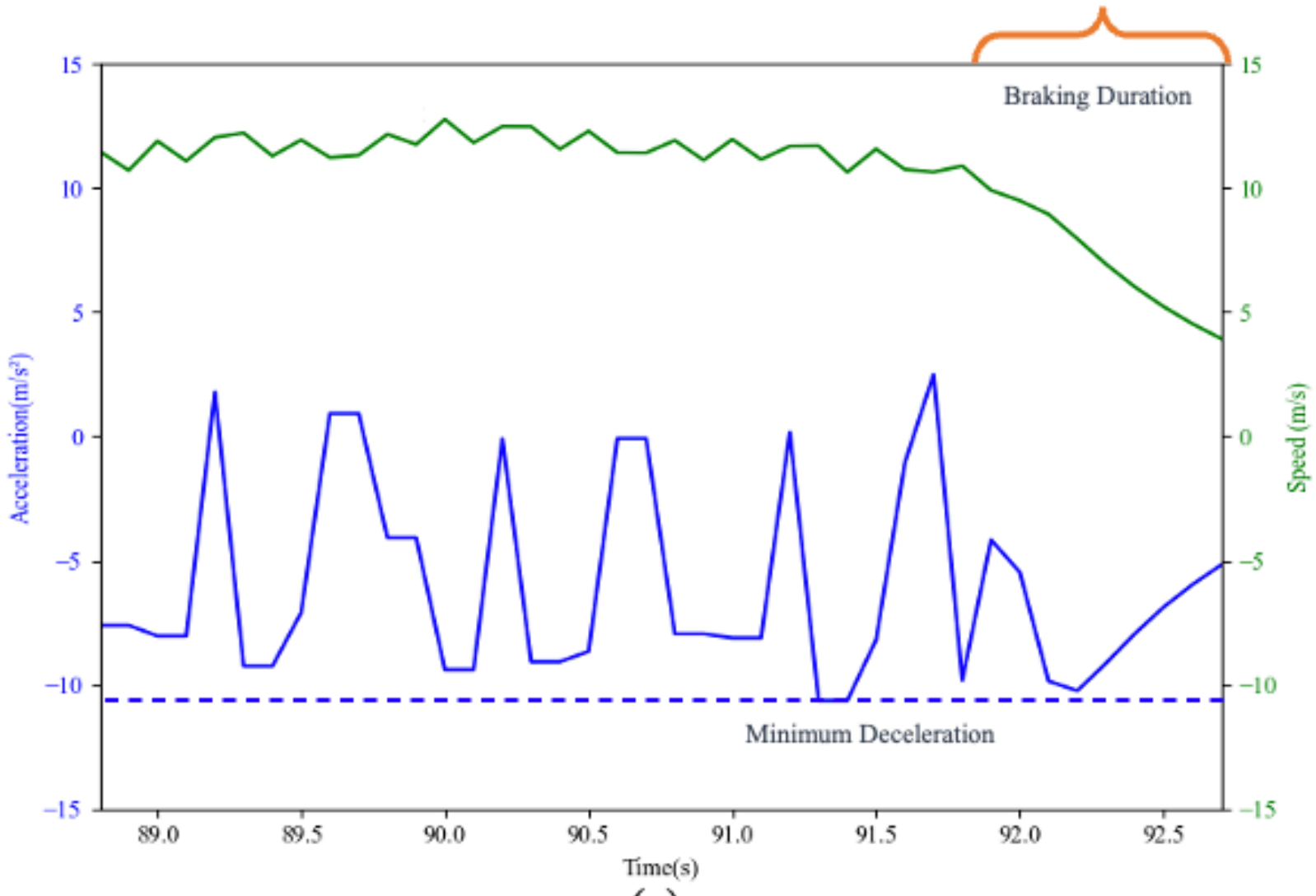
Braking Duration
Minimum Deceleration
Acceleration(m/s²)
Speed (m/s)
Time(s)
89.0 89.5 90.0 90.5 91.0 91.5 92.0 92.5

(a)

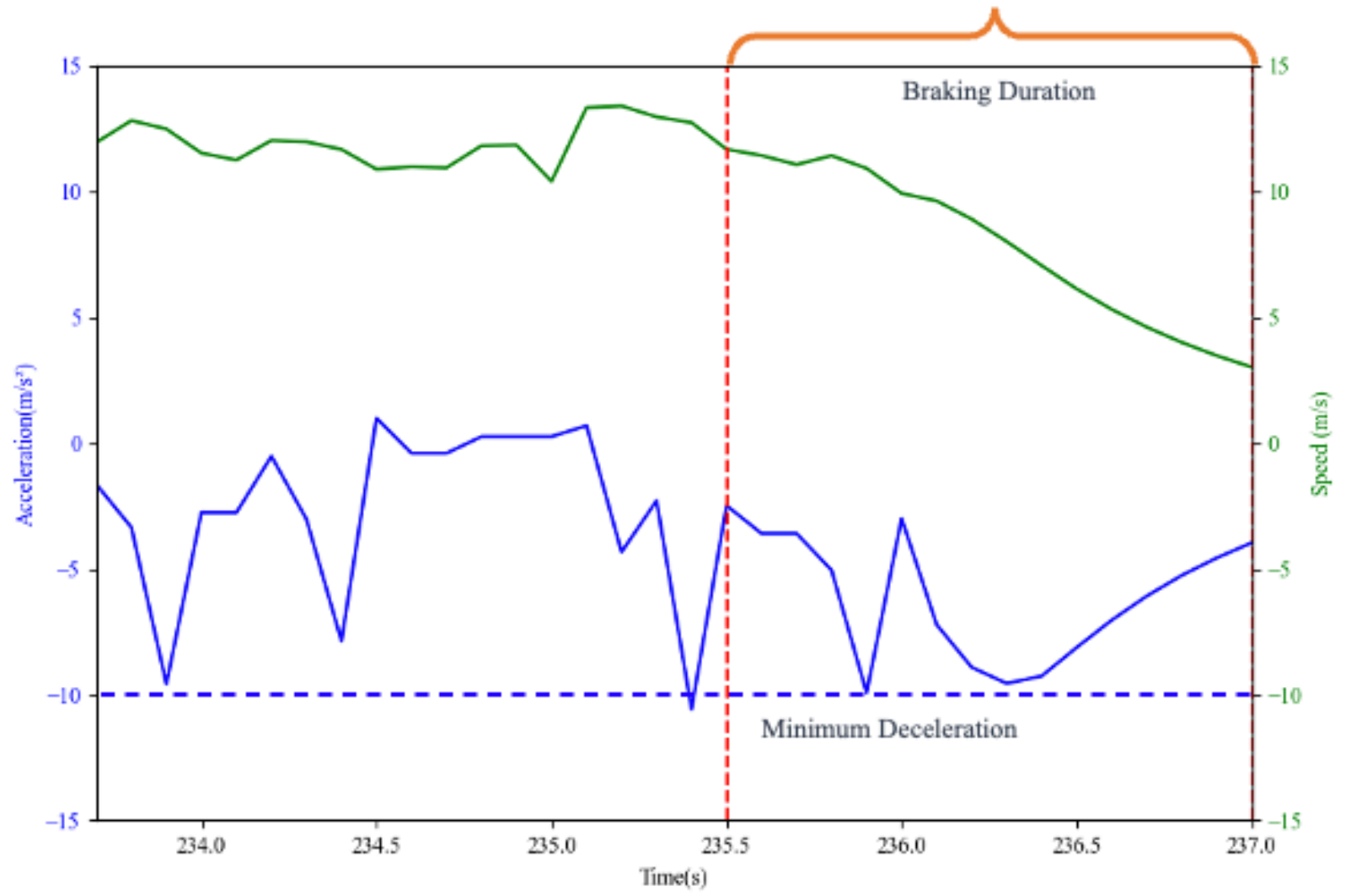

(b)

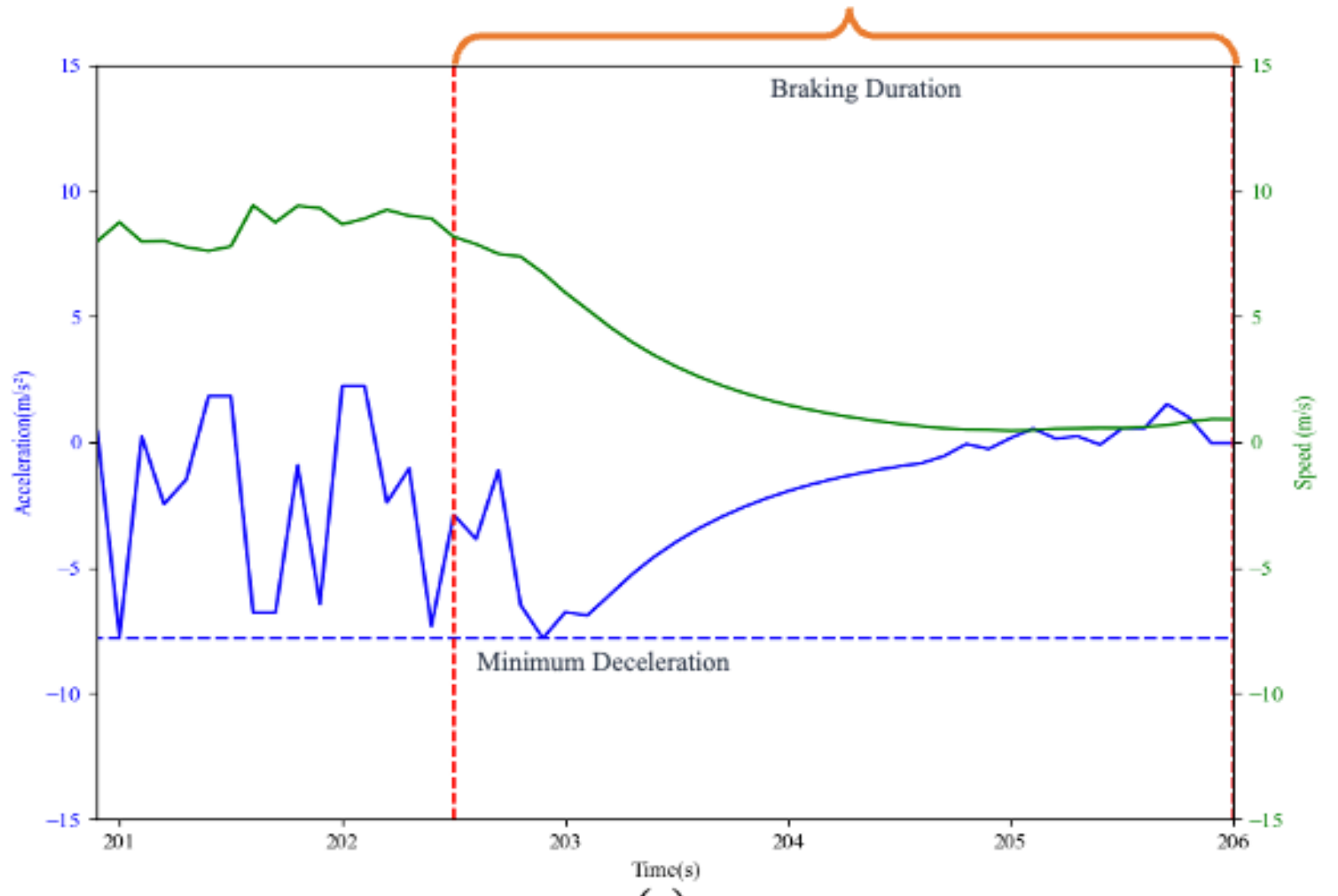

(c)

**Figure 7** Speed and acceleration profile comparison with no warning (a), warning issued 1 second in advance (b), warning issued 2 seconds in advance (c)

For the impact of the aggressive level, TTC, CPI, and DRAC show similar statistical characteristics. Specifically, the low aggressive level resulted in the highest TTC ($F$(1,35) =10.358, $p$ = .003), lowest CPI ($F$(1,35)=19.059, $p$<.001)) and highest DRAC ($F$(1,35)=54.618, $p$=.001) compared to the other two aggressive levels. These findings aligned with our hypothesis that lower aggression levels lead to safer traffic conditions. As shown in **Figure 7(e)**, the high aggressive level had a higher maximum DRAC than the medium aggressive level ($F$(1,35)=6.917, $p$=.013). However, there was no significant difference in TTC between the medium and high aggressive levels (as shown in **Figure 6(b)**), and CPI even showed that the CPI in the medium aggressive level was higher ($F$(1,35)=9.214, $p$=.005) than the high aggressive level as shown in **Figure 6(d)**. This pattern is counterintuitive because usually a higher aggressive level results in a lower (higher) TTC (CPI). To explain these data, our hypothesis is that under high aggressive level scenarios, the drivers perceived higher risks and thus took more aggressive actions (i.e., deceleration) to prevent potential crashes. To validate this hypothesis, we analyzed the impact of aggressive levels and warnings on the risk perception as shown in **Figure 8** and **Table 2**.

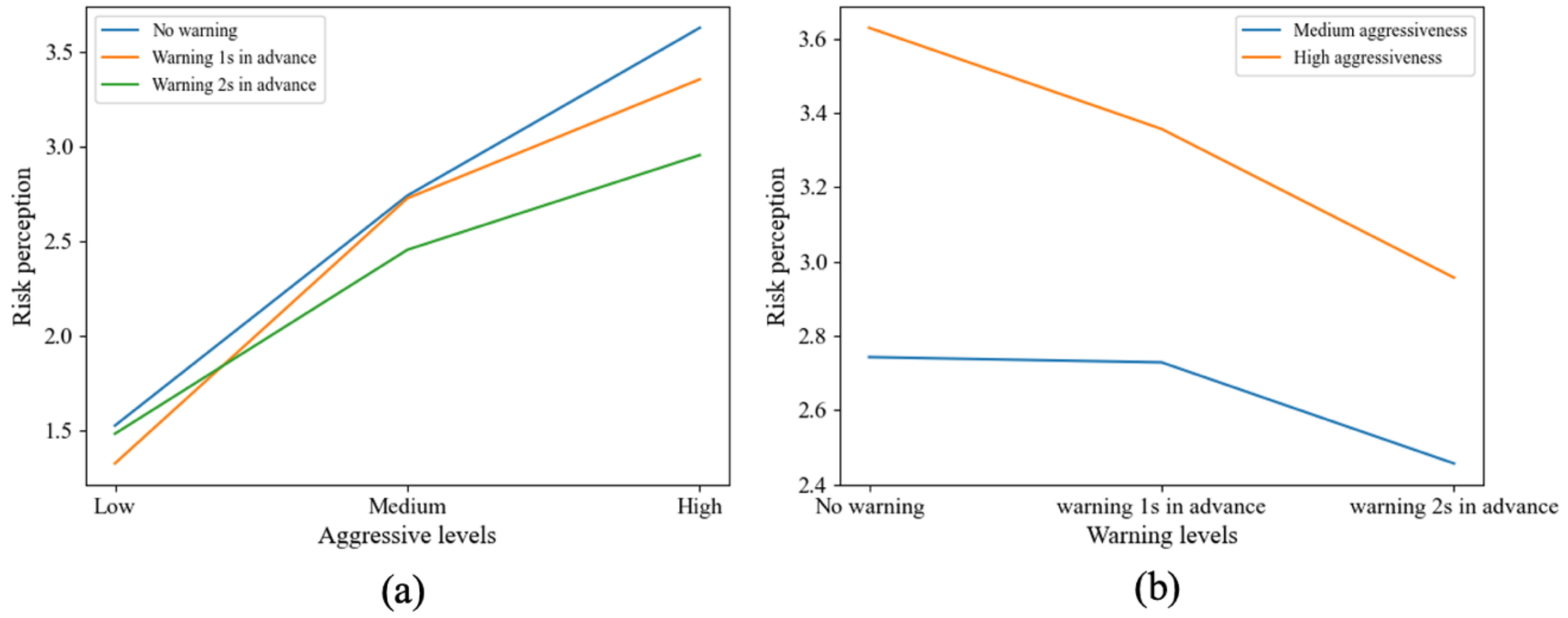

**Figure 8** Risk perception by aggressive levels (a) and warning levels (b)

**Table 2 Results of ANOVA on risk perception to investigate the effect of warning and aggressiveness**

| Factors | df | Mean Square | F | $p$ | Partial Eta Squared |
|---|---|---|---|---|---|
| Warning | 2 | 3.968 | 5.341 | 0.007 | 0.132 |
| Aggressiveness | 2 | 92.26 | 97.471 | <.001 | 0.736 |
| Warning * Aggressiveness | 4 | 1.644 | 2.86 | 0.026 | 0.076 |

The ANOVA indicated that the low aggressive level led to lower risk perception than the other two levels ($F$(1, 35)=158.484, $p$<.001). The medium aggressive level also resulted in lower risk perception than the high aggressive level ($F$(1, 35)=28.209, $p$ <.001), as shown in Figure 9(a). Meanwhile, no warning scenarios resulted in the highest risk perception compared with the other two conditions ($F$(1,35)=4.789, $p$=.035). Additionally, warnings issued 2 seconds in advance led to lower risk perception than the other two conditions ($F$(1,35)=6.136, $p$=.018), as shown in **Figure 8(b)**. In summary, the survey showed that drivers felt higher risk perception on the high aggressive level compared with the medium aggressive level, but counterintuitive results of TTC and CPI were found.

Given that the drivers perceived higher risks in more aggressive scenarios, we next examined the braking behavior to explain similar TTCs. In the experiment, we noticed that different drivers employed different crash avoidance strategies. Some drivers opted for immediate avoidance and abruptly slammed on the brakes when they noticed the warning, while others adopted a more gradual and mild braking strategy over a longer distance. Consequently, solely examining key time points (like minimum TTC) is insufficient to understand the braking behavior. As a result, we considered the time interval $[-2, min(+2,)]$ as the "braking period" to extract data (e.g., average deceleration rate and maximum deceleration rate within the "braking period") to describe braking behaviors. is the time point at which the vehicle arrives at the collision point.

Drivers had a higher average deceleration ($F$(1,35)=36.119, $p$<.001) in the high aggressive level compared to the medium aggressive level, as shown in **Figure 9(a),** and a higher maximum deceleration rate ($F$(1,35)=7.391, $p$=0.01), as shown in **Figure 9(b)**. The analyses from the risk perception and braking behavior validate our hypothesis that under higher aggressive levels, the drivers perceived higher risks and applied higher deceleration, which results in similar TTCs.

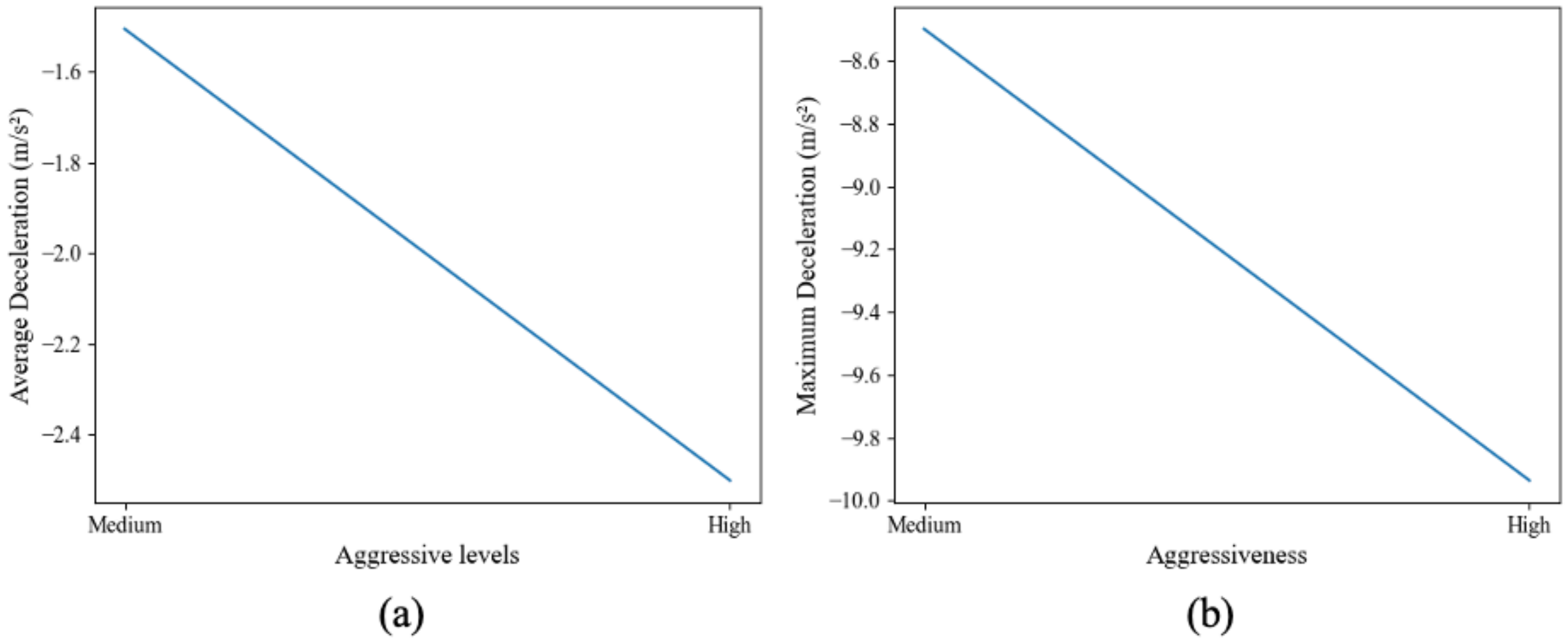

**Figure 9** The braking behavior by aggressive levels. (a) Average deceleration (b) Maximum deceleration

## STOP-OR-GO DECISION PREDICTION MODEL

In the experiment, we observed that in 24.65% ($N$=71) of the cases, the driver disregarded the warning and attempted to accelerate and rush out before the aggressive vehicle. The "rush out" cases were significantly more dangerous as they have a much lower minimum TTC compared to the cases where they stop for the aggressive vehicle ("rash out" TTC= 0.282s, "stop" TTC=0.909s, welch' *t*=-7.024, *p*<.001). As the "rush out" cases exhibited very dangerous driving situations due to extremely low minimum TTCs, countermeasures must be taken to prevent such behaviors. As the first step, the driver's intention needs to be predicted (i.e., whether to follow the warning and yield to the aggressive vehicle). This motivated us to develop a personalized intention prediction model to predict the driver's stop-or-go decision under warning conditions.

The prediction was conducted precisely at the warning start time, focusing on discerning the driver's intent of "stop or go" decision under the influence of the warning. Six features were selected for the prediction model. These features can be categorized into three types: vehicle trajectory features (e.g., speed of ego vehicle $v_i$, and time headway to the conflict point $h_t$), safety features (e.g., acceleration noise AN, deceleration rate to avoid the crash DRAC), and eye tracker features (e.g., mean fixation duration on road $MFD_{road}$, and average pupil diameter $\overline{PD}$). Among these, $v_i$, $h_t$, and DRAC are instantaneous metrics measured at the warning start time point, while AN, $MFD_{road}$ and $\overline{PD}$ are the metrics encompassing the time from the moment of entering the roundabout to the onset of the warning. All six features can be collected in real time from the ego vehicle. Some features are directly related to a specific driver's driving behavior (e.g., AN and DRAC) and eye movements. As a result, we considered this to be a personalized prediction model.

The Pearson correlation coefficient matrix was utilized to evaluate the correlation between these features to avoid collinearity. Among the variable pairs examined, three pairs of variables (e.g., speed and headway, speed and AN, headway and AN) have low correlations (0.3-0.5), while the others display negligible correlations (less than 0.3). Several machine learning (ML) models were employed to predict the drivers' intent, including the K-Nearest Neighbors (KNN), decision tree, random forest, and XGboost. A total of 288 data points were applied to the prediction model. 80% of the data was utilized for training and 20% for testing. **Table 3** presents the prediction results obtained from these models. Among them, XGboost demonstrates the highest prediction accuracy of 94.8%. Overall, all ML models exhibit good performance in predicting drivers' intent, as evidenced by high precision, recall, and F1 score values. The precision values, representing the accuracy of positive predictions, range from

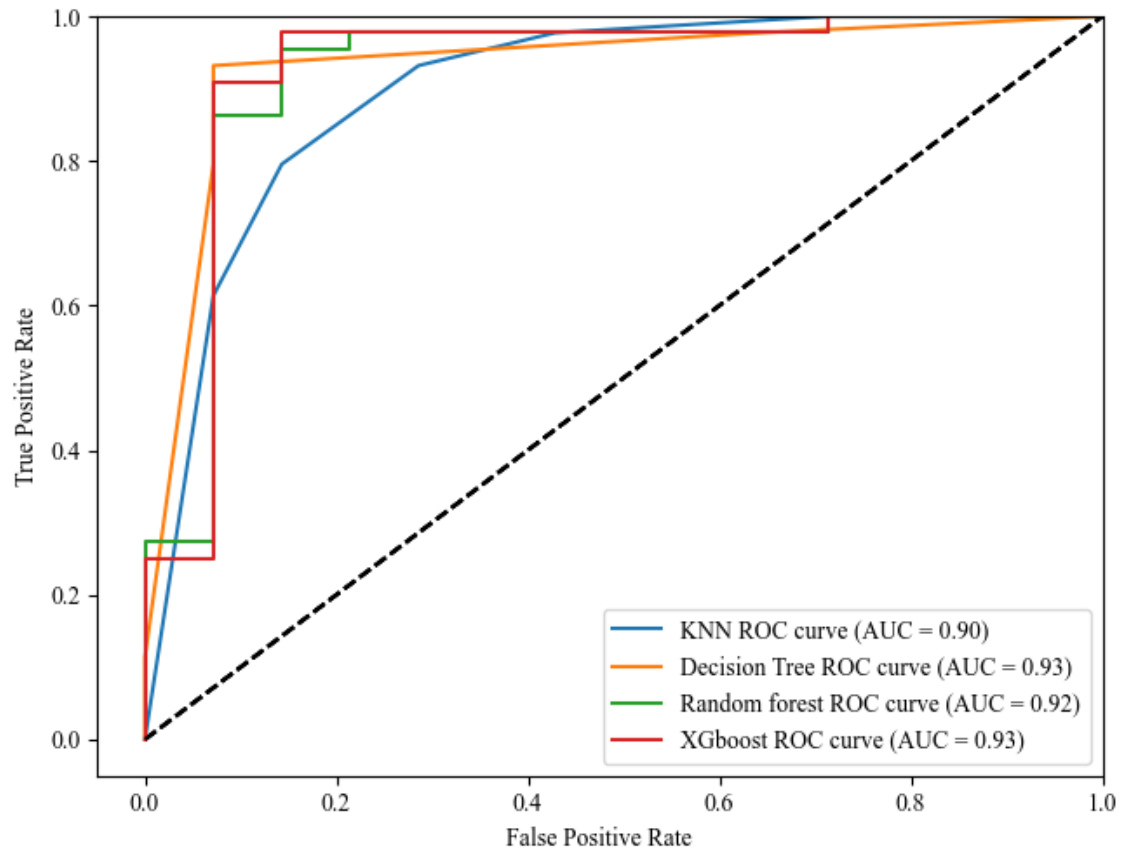

0.911 to 0.976. The recall values, which indicated the proportion of correctly identified positive cases, range from 0.909 to 0.977 across different ML models. The F1 scores, which combined precision and recall range from 0.921 to 0.966. **Figure 10** shows the ROC curves of each ML model. The decision tree and XGboost have the best performance with an AUC value of up to 93%.

**Figure 10 ROC curve of various ML prediction models**

**Table 3 ML models prediction results**

| Classifier | Training accuracy | Testing accuracy | Precision | Recall | F1 score |
|---|---|---|---|---|---|
| KNN | 0.874 | 0.879 | 0.911 | 0.932 | 0.921 |
| Decision tree | 0.917 | 0.914 | 0.976 | 0.909 | 0.941 |
| Random forest | 0.935 | 0.931 | 0.955 | 0.955 | 0.955 |
| XGboost | 0.957 | 0.948 | 0.956 | 0.977 | 0.966 |

In this section, we aim to gain insights into the driver's decision-making process in the presence of the warning, thereby enhancing our understanding of what driving features are related to hazardous driving behaviors. If the prediction model indicates that the driver would ignore the warning and choose to "rush out", further countermeasures should be taken. For example, an elevated warning can be issued through other forms such as audio or vibration of the steering wheel. On the other hand, the vehicle could apply automatic emergency brakes, which is now a standard function in many commercial vehicles that are equipped with ADAS. Details of the countermeasure design are beyond the scope of this paper and will be investigated in further studies.

## CONCLUSIONS

We conducted a study to validate an infrastructure-based warning system designed to improve roundabout driving safety. To create a realistic and flexible testing environment, we developed a co-simulation platform that combines Webots for vehicle/driving simulation and SUMO for traffic simulation. This co-simulation platform facilitates the integration of real-world maps, precise control of vehicle states, generation of diverse traffic conditions, and seamless incorporation of V2X technology with sensing and information-processing capabilities.

A driving simulator experiment was conducted to investigate how the infrastructure-based warning system influences drivers' behavior and decision-making. We analyzed the effects of warnings and aggressiveness using ANOVA. The results demonstrate that the presence of warnings significantly contributed to maintaining a safer driving environment compared to scenarios without warnings. Furthermore, earlier warnings proved particularly beneficial as they allowed drivers to react in a longer time and with greater comfort and smoothness, reducing the occurrence of sudden and harsh decelerations These results emphasized the effectiveness of the infrastructure-based warning system in enhancing driving safety and reducing the likelihood of accidents within a roundabout.

In addition, personalized intent prediction models based on several ML techniques were developed to predict the drivers' stop-or-go decisions when the warning onset. Among all ML models, XGBoost achieved a remarkable prediction accuracy with 97.73% recall and 95.56% precision. The prediction model serves as a foundation for designing personalized advanced warning systems that are customized to different drivers in future research.

There are a few future research directions. First, the behavior of the aggressive vehicle is predefined in the present study. An accurate prediction model is needed in real-world implementation to determine whether the vehicle will yield or not. Second, the co-simulation platform can be further extended to support experiments between human-driven vehicles (HDVs) and connected and automated vehicles (CAVs) by integrating the autonomous driving stack (e.g., Autoware) into the Webots vehicles. Another promising direction worth exploring is the design of further countermeasures aiming at effectively alerting drivers who tend to ignore the warnings.

## ACKNOWLEDGEMENT

This research is supported in part by the U.S. National Science Foundation (NSF) through Grant No. 2121967. The views presented in this paper are those of the authors alone.

## AUTHOR CONTRIBUTION STATEMENT

Study conception and design: Zhang, Tian, Han, Feng, Chen, Proctor, Zhang; Analysis and interpretation of results: Zhang, Tian, Han, Li, Feng, Chen, Proctor, Zhang; Manuscript preparation: All authors; Funding acquisition: Feng, Chen, Proctor, Zhang.